\documentclass[useAMS,usenatbib]{mn2e}

\usepackage[T1]{fontenc}
\usepackage{pslatex}
\usepackage{amsmath}
\usepackage{amsfonts}
\usepackage{color}
\usepackage{url}
\usepackage{graphicx}
\usepackage{subfigure}
\usepackage{amssymb}
\usepackage{textcomp}
\usepackage{soul}
\usepackage{booktabs}
\usepackage{array}
\usepackage{hyperref}
\usepackage{enumerate}
 \graphicspath{{images/}}

{\newif\ifnotend
\notendtrue
\def\veclist{ABCDEFGHIJKLMNOPQRSTUVWXYZabcdefghijklmnopqrstuvwxyz.}
\def\top#1#2.{#1}
\def\tail#1#2.{#2.}
\loop\expandafter\xdef\csname v\expandafter\top\veclist\endcsname%
{{\noexpand\bf\expandafter\top\veclist}}
\edef\veclist{\expandafter\tail\veclist}
\if\veclist.\notendfalse\fi\ifnotend\repeat}
\def\pa{\partial}

\DeclareMathOperator{\sech}{sech}

\mathchardef\mhyphen="2D

\title[The geometry of gas surrounding the CMZ]{The geometry of the gas surrounding the Central Molecular Zone: on the origin of localised molecular clouds with extreme velocity dispersions}
\author[Sormani et al.]{Mattia C. Sormani$^{1}$, Robin G. Tre{\ss}$^1$, Simon C.O. Glover$^1$, Ralf S. Klessen$^{1,2}$,  \newauthor Ashley T. Barnes$^3$, Cara D. Battersby$^4$, Paul C. Clark$^5$, H Perry Hatchfield$^4$ \newauthor and Rowan J. Smith$^6$\\
$^1$Universit\"{a}t Heidelberg, Zentrum f\"{u}r Astronomie, Institut f\"{u}r theoretische Astrophysik, Albert-Ueberle-Str. 2, 69120 Heidelberg, Germany \\
$^2$Universit\"at Heidelberg, Interdiszipli\"ares Zentrum f\"ur Wissenschaftliches Rechnen, Im Neuenheimer Feld 205, 69120 Heidelberg, Germany \\
$^3$Argelander Institute for Astronomy, University of Bonn, Auf dem H\"ugel 71, D-53121 Bonn, Germany \\
$^4$University of Connecticut, Department of Physics, 2152 Hillside Road,U-3046, Storrs, CT 06269, USA \\
$^5$School of Physics and Astronomy, Queen's Buildings, The Parade, Cardiff University, Cardiff, CF24 3AA, UK \\
$^6$Jodrell Bank Centre for Astrophysics, School of Physics and Astronomy, University of Manchester, Oxford Road, Manchester M13 9PL, UK \\
}

\begin{document}
\hyphenation{kruijs-sen}

\date{}

\def\p{\partial}
\def\Omegap{\Omega_{\rm p}}

\newcommand{\di}{\mathrm{d}}
\newcommand{\bfx}{\mathbf{x}}
\newcommand{\bfe}{\mathbf{e}}
\newcommand{\vlos}{\mathrm{v}_{\rm los}}
\newcommand{\Tspin}{T_{\rm s}}
\newcommand{\Tb}{T_{\rm b}}
\newcommand{\degree}{\ensuremath{^\circ}}
\newcommand{\Th}{T_{\rm h}}
\newcommand{\Tc}{T_{\rm c}}
\newcommand{\bfr}{\mathbf{r}}
\newcommand{\bfv}{\mathbf{v}}
\newcommand{\pc}{\,{\rm pc}}
\newcommand{\kpc}{\,{\rm kpc}}
\newcommand{\Myr}{\,{\rm Myr}}
\newcommand{\Gyr}{\,{\rm Gyr}}
\newcommand{\kms}{\,{\rm km\, s^{-1}}}
\newcommand{\de}[2]{\frac{\partial #1}{\partial {#2}}}
\newcommand{\cs}{c_{\rm s}}
\newcommand{\rb}{r_{\rm b}}
\newcommand{\rqu}{r_{\rm q}}
\newcommand{\nuP}{\nu_{\rm P}}
\newcommand{\thetaobs}{\theta_{\rm obs}}
\newcommand{\hatn}{\hat{\textbf{n}}}
\newcommand{\hatx}{\hat{\textbf{x}}}
\newcommand{\haty}{\hat{\textbf{y}}}
\newcommand{\hatz}{\hat{\textbf{z}}}
\newcommand{\hatX}{\hat{\textbf{X}}}
\newcommand{\hatY}{\hat{\textbf{Y}}}
\newcommand{\hatZ}{\hat{\textbf{Z}}}
\newcommand{\hatN}{\hat{\textbf{N}}}

\maketitle

\begin{abstract}
Observations of molecular gas near the Galactic centre ($| l | < 10 \degree$, $| b | < 1 \degree$) reveal the presence of a distinct population of enigmatic compact clouds which are characterised by extreme velocity dispersions ($\Delta v > 100\kms$). These Extended Velocity Features (EVFs) are very prominent in the datacubes and dominate the kinematics of molecular gas just outside the Central Molecular Zone (CMZ). The prototypical example of such a cloud is Bania Clump 2. We show that similar features are naturally produced in simulations of gas flow in a realistic barred potential. We analyse the structure of the features obtained in the simulations and use this to interpret the observations. We find that the features arise from collisions between material that has been infalling rapidly along the dust lanes of the Milky Way bar and material that belongs to one of the following two categories: (i) material that has `overshot' after falling down the dust lanes on the opposite side; (ii) material which is part of the CMZ. Both types of collisions involve gas with large differences in the line-of-sight velocities, which is what produces the observed extreme velocity dispersions. Examples of both categories can be identified in the observations. If our interpretation is correct, we are directly witnessing (a) collisions of clouds with relative speeds of $\sim 200\kms$ and (b) the process of accretion of fresh gas onto the CMZ.
\end{abstract}

%%%
%%%

\section{Introduction}
\label{sec:intro}

The geometry of the gas in the Central Molecular Zone (CMZ -- defined here as the region at a radial distance $R\lesssim 200\pc$ from the Galactic centre, or equivalently $ | l | \lesssim 1.5 \degree$) has been intensively studied in recent years \citep[e.g.][]{Molinari+2011,Kruijssen+2015,Henshaw+16a,Sormani+2018}. In contrast, the region immediately surrounding the CMZ has received relatively little attention. However, it is well known that the CMZ is not an isolated system, but instead is strongly interacting with its surroundings. For example, the Galactic bar continuously drives a gas inflow into the CMZ which strongly affects its dynamics and may even drive the observed turbulence of the CMZ \citep{SormaniBarnes2019}.

Among the most enigmatic features in the region surrounding the CMZ is a discrete population of extremely broad-lined ($\Delta v > 100 \kms$) compact clouds that are very prominent in molecular line datacubes (e.g.\ CO) in the region $| l |\leq 10 \degree$ \citep[][]{Liszt2006,Liszt2008}. These features dominate the kinematics of molecular gas just outside the CMZ. The prototypical example is Bania Clump 2 \citep{StarkBania1986}. Despite their enormous velocity dispersion, these puzzling features are confined to a narrow longitude range. Similar features are not found anywhere else in the Galaxy. In this paper, we will refer to these features as Extended Velocity Features (EVFs) on account of their large velocity dispersions. We give a brief summary of the observational properties of the EVFs in Sect. \ref{sec:observations}. 

Several possible interpretations of the EVFs have been put forward in the literature:
\begin{enumerate}[(a)]
\item They are gaseous structures extended in space that happen to coincidentally lie parallel to the line of sight \citep[e.g.][]{StarkBania1986,Boyce+89,Lee++1999,Baba++2010}. Such interpretations suffer from the `fingers of god' effect, i.e. they assume that we are at a special location in the universe in which these structures happen to point toward us.
\item Some of them have been interpreted as the footprints of giant magnetic loops caused by the \cite{Parker1966} instability near the Galactic centre \citep{Fukui+2006,Fujishita+2009,Machida+2009,Torii+2010,Suzuki+2015,Riquelme+2018}.
\item Some of them have been interpreted as evidence for the presence of intermediate-mass black holes (IMBH) \citep{Oka+2016,Oka+2017,Takekawa+2019,Takekawa+2019b}.
\item They are lumps which are just about to cross the dust lanes of the Milky Way bar (\citealt{Fux1999}, see also \citealt{Liszt2006,Liszt2008}).
\end{enumerate}

In this paper we show that features similar to the observed ones arise naturally in simulations of gas flow in a barred potential. We then use the insight gained from the simulations to interpret the observations. The paper is structured as follows. In Sect. \ref{sec:observations} we briefly review the observations and the key properties that characterise the EVFs. In Sect. \ref{sec:setup} we describe the numerical setup of our simulations. In Sects. \ref{sec:results} and \ref{sec:discussion} we discuss our results and interpret the observations. Finally in Sect. \ref{sec:conclusion} we sum up.

\section{Observations}
\label{sec:observations}

Here we briefly review the observational data. A more detailed analysis can be found for example in \citet{Liszt2006,Liszt2008} and \citet{Oka+2012} for CO, \citet{BoyceCohen1994} for OH, \citet{Longmore+2017} for NH$_3$ and \citet{McClureGriffiths+2012} for HI.

\begin{figure*}
\includegraphics[width=0.8\textwidth]{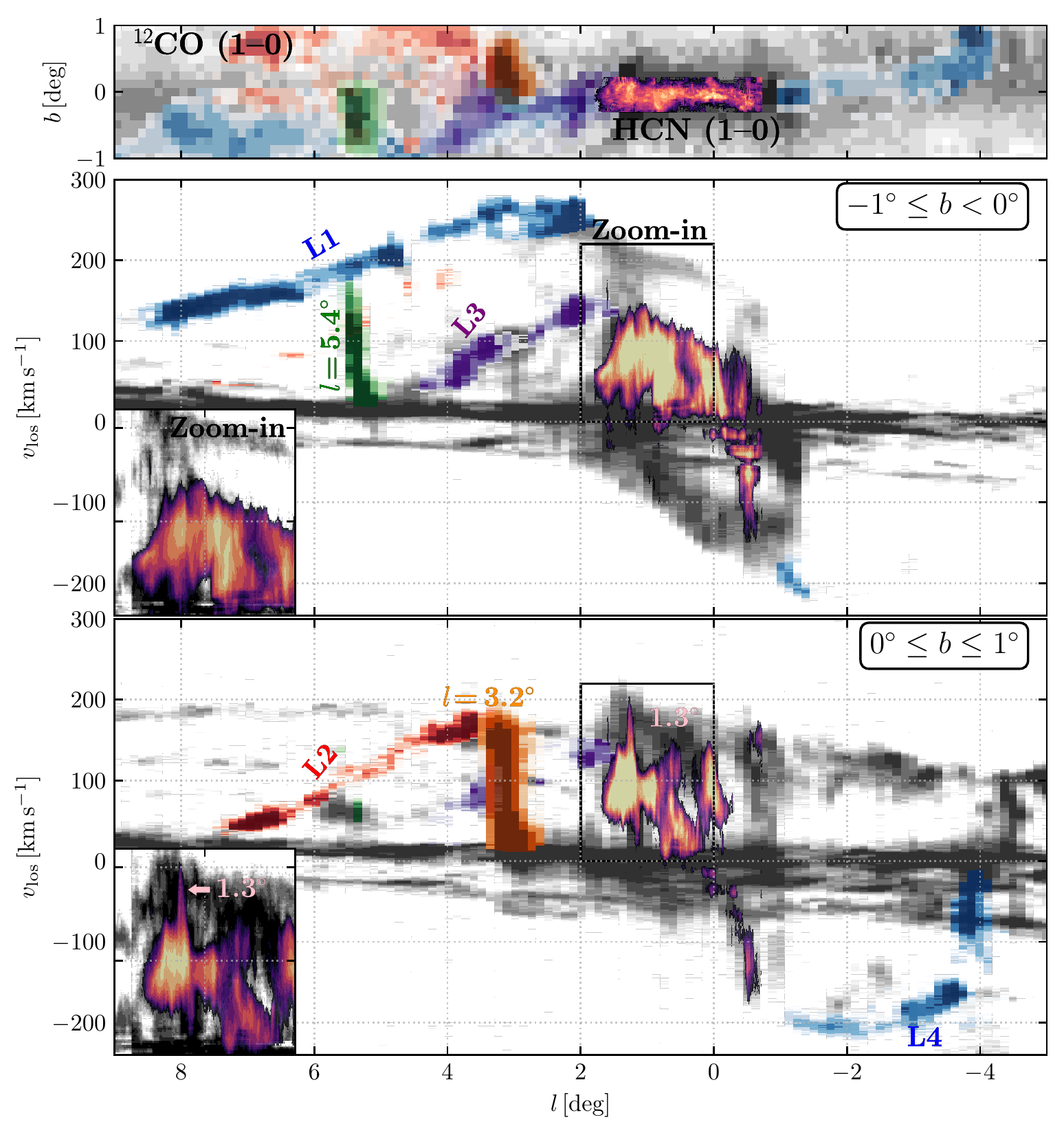}
\caption{Molecular emission from the inner Galaxy. Some of the most prominent EVFs ($l=1.3\degree$, $l=3.2\degree$ a.k.a. Bania Clump 2 and $l=5.4\degree$) and the dustlane-like features identified by \citet{Liszt2008} (L1 to L4) are indicated. The grey background shows the $^{12}$CO $J=1\to0$ data from \citet{Bitran+1997} (in the main panels) and \citet{Oka+1998} (in the zoom-in panels). The $l=5.4\degree$, $l=3.2\degree$ and L1 to L4 features are highlighted in the CO data. The magma colour scale in the centre shows HCN from the data of \citet{Jones+2012}. The HCN data only covers the region $ -0.7 < l < 1.8  \degree$, $-0.3<b< 0.2\degree$ and $-300 <v< 300 \kms$. The $l=1.3\degree$ feature is visible in the HCN data and is indicated with an arrow.}
\label{fig:data}
\end{figure*}

Fig. \ref{fig:data} shows molecular line emission from the inner Galaxy. The three most prominent EVFs are highlighted: these are the $l=5.4\degree$, the $l=3.2\degree$ (also known as Bania Clump 2) and the $l=1.3\degree$ features. Other, less prominent EVFs can be found throughout the inner regions of our Galaxy (see references above). 

Also highlighted are the dust lane features L1 to L4. These are not EVFs, but are often linked to them in $(l,b,v)$ space (see Property vi below). The "dust lane" terminology is used here for historical reasons despite these features being primarily (but not exclusively) detected in gas. The terminology originally comes from observations of external barred galaxies such as NGC 1300 or NGC 5383 in which one can see "the presence of two dust lanes leaving the nucleus one on each side of the bar and extending into the spiral arms" \citep{Sandage1961}. After it was realised that the MW is a barred galaxy, the features L1 and L4 were identified as the dust lanes of the MW bar \citep{Fux1999}, and the ``dust lane'' terminology was maintained despite the fact that they were initially observed in HI and CO emission, and not from dust emission/extinction. Subsequent work has identified the L1 and L4 features also from the dust \citep{Marshall+2008}. Beyond the two main dust lane features L1 and L4, \cite{Liszt2008} determined the presence of the two additional secondary dust lane features L2 and L3 using CO emission. As we will see later in the paper, the presence of multiple dust lanes also occurs in our simulations.

The key properties that characterise the EVFs are:
\begin{enumerate}
\item They are extremely broad-lined, with velocity dispersions of up to $200 \kms$ when observed at low resolution.
\item They are compact, so they are very localised in the $(l,b)$ plane (the typical extensions of the largest EVFs are $\Delta l, \Delta b \sim 0.5\degree$, )
\item They are usually more extended in latitude than in longitude. So they are typically elongated perpendicularly to the Galactic plane.
\item They are predominantly found in the $(v>0,l>0)$ and $(v<0,l<0)$ quadrants of the $(l,v)$ plane, although a few of them are found in the other two quadrants as well.
\item They never extend beyond the Terminal Velocity Curve (TVC)\footnote{The TVC at $l>0$ ($l<0$) is defined as the maximum (minimum) value of line-of-sight velocity at which the bulk of the emission from the Galactic disc is found, i.e. it is the curve that defines the envelope of the latitude-integrated $(l,v)$ diagram \citep[see for example][Chapter 9]{BM}.} at their value of longitude.
\item Some of them are clearly connected to some dustlane-like features associated with the Galactic bar (see for example how the $l=5.4\degree$ EVF connects L1 to L3 or how the $l=3.2\degree$ EVF is connected to L2, see also \citealt{Liszt2008}).
\item Some of them (e.g. Bania's Clump 2) show sharp HI emission profiles on one side \citep{McClureGriffiths+2012}.
\item When observed at very high resolution, they typically break-up into multiple kinematic sub-components with strong velocity gradients (see for example \citealt{Liszt2006} which resolved the internal velocity structure of several EVFs and fig. 30 in \citealt{Longmore+2017} which shows the complicated velocity structure of Bania Clump 2 in NH$_3$).
\end{enumerate}

Successful theoretical models should be able to reproduce the above properties.

\section{Numerical setup} \label{sec:setup}

Our numerical setup is the same as that of \cite{Sormani+2018} except for a few differences. Therefore we only provide here a brief recap and state the differences from these previous simulations, and refer the reader to section 3 of \cite{Sormani+2018} and references therein for a more detailed description.

\subsection{Hydrodynamic code}

The simulations are three-dimensional and the gas is assumed to flow in a multi-component external barred potential $\Phi(\bfx,t)$ which is constructed to fit the properties of the Milky Way (see next section and Appendix \ref{sec:potential}). The gas self-gravity and magnetic fields are neglected.

We use the moving-mesh code {\sc Arepo} \citep{Springel2010}, modified to treat the chemical evolution of the interstellar gas. The code solves the equations of fluid dynamics:
\begin{align}
& \frac{\pa \rho}{\pa t} + \nabla \cdot (\rho \bfv) = 0, \\
& \frac{ \pa (\rho \bfv)}{\pa t} + \nabla \cdot \left( \rho \bfv \otimes \bfv + P \bold I \right) = - \rho \nabla \Phi, \\
& \frac{\pa(\rho e)}{\pa t} + \nabla \cdot \left[(\rho e + P) \bfv \right] = \dot{Q} + \rho \frac{\pa \Phi}{\pa t},  \label{eq:energy}
\end{align}
where $\rho$ is the gas density, $\bfv$ is the velocity, $P$ is the thermal pressure, $\bold I$ is the identity matrix, $e~=~e_{\rm therm} + \Phi + {\bfv^2}/{2} $ is the energy per unit mass, $e_{\rm therm}$ is the thermal energy per unit mass. We adopt the equation of state of an ideal gas, $P = (\gamma -1) \rho e_{\rm therm}$, where $\gamma=5/3$ is the adiabatic index. 

We account for the chemical evolution of the gas using an updated version of the NL97 chemical network from \citet{gc12}, which itself was based on the work of  \citet{gm07a,gm07b} and \citet{nl97}. With this network, we solve for the non-equilibrium abundances of H, H$_{2}$, H$^{+}$, C$^{+}$, O, CO and free electrons. An extensive description of the network is given in Section~3.4 of \citet{Sormani+2018} and in the interests of brevity we do not repeat it here.

The term $\dot{Q}$ in Equation~\ref{eq:energy} contains the contributions of the radiative and chemical processes that can change the internal energy of the system ($\dot{Q}=0$ for an adiabatic gas). It includes (i) a cooling function which depends on the instantaneous chemical composition of the gas \citep{glo10,gc12}; (ii) the heat absorbed or released in the most important chemical processes that occur in the interstellar medium, which are tracked in real time by the chemical network; (iii) external heating sources that represent the average Interstellar Radiation Field (ISRF) and cosmic ray ionisation rate. The strength of the ISRF is set to the standard value $G_0$ measured in the Solar neighbourhood \citep{draine78} diminished by a local attenuation factor which depends on the amount of gas present within 30~pc of each computational cell. This attenuation factor is introduced to account for the effects of dust extinction and H$_{2}$ self-shielding and is calculated using the {\sc Treecol} algorithm described in \cite{clark12}. The cosmic ray ionisation rate is fixed to $\zeta_{\rm H} = 3 \times 10^{-17} \: {\rm s^{-1}}$ \citep{gl78}. These values correspond to the `low' simulation of \cite{Sormani+2018}. We have shown in that paper the strength of the ISRF mainly controls the amount of molecular gas but makes little difference to the dynamics. Indeed, even if the ISRF field is a factor of a 1000 higher than in the Solar Neighbourhood \citep{Clark+2013}, the sound speed of the molecular gas comes nowhere close to the values of $\cs=5\mhyphen10\kms$ which would be needed to significantly affect the dynamics of the gas \citep{SBM2015a}. Hence the results of the present paper are not affected by the strength of the assumed ISRF. 

\subsection{Differences between \citet{Sormani+2018} and the present paper}

The main difference between the simulations in \cite{Sormani+2018} and the one used in the present paper is that we modified the gravitational potential of the bar so that the size of the nuclear ring that naturally forms in the simulation matches the observed size of the CMZ (it was a factor of $\sim 2$ too large in the previous simulations). In general, the size of this ring is controlled by (i) the parameters of the gravitational potential, mainly the bar strength and the bar pattern speed \citep[e.g.][and references therein]{Sormani+2018b} (ii) the effective sound speed of the gas (see for example Fig.~1 of \citealt{SBM2015a}). Since the sound speed of the gas is fixed by our treatment of the heating \& cooling of the ISM and the pattern speed of the gas is independently constrained to be $\Omega_{\rm p} = 40 \kms \kpc^{-1}$ \citep[e.g.][]{SBM2015c,Portail+2017,Sanders+2019}, we have increased the strength of the bar (compatibly with with known observational constraints) to achieve the desired result of a smaller ring. The gravitational potential and the resulting rotation curve are described in detail in Appendix \ref{sec:potential}.

The second difference is that we increased the resolution. The resolution in the simulation is determined by the condition that cells approximately have the same mass (so that denser gas has a higher spatial resolution). The system of mass refinement present in {\sc Arepo} ensures that this condition is satisfied by splitting cells whose mass becomes greater than twice this target mass and merging cells whose mass is too low. Here we use a target resolution of  $25 \, \rm M_\odot$ per cell, while in \cite{Sormani+2018} we used a target resolution of  $100 \, \rm M_\odot$.

The last difference is in the initial density profile of the gas. In \cite{Sormani+2018} the initial density distribution was approximately uniform inside a cylindrical slab of radius $10~\kpc$ and half-height $1~\kpc$, with the addition of some small random noise. Here instead we initialise the density according to the following axisymmetric density distribution:
\begin{equation}
\rho(R,z) = \frac{\Sigma_0}{4 z_{\rm d}}  \exp\left(- \frac{R_{\rm m}}{R} - \frac{R}{R_{\rm d}}\right) \sech\left(\frac{z}{2 z_{\rm d}}\right)^2\, ,
\end{equation}
where $(R,\phi,z)$ denote standard cylindrical coordinates, $z_{\rm d} = 85 \pc$, $R_{\rm d} = 7 \kpc$, $R_{\rm m} = 1.5 \kpc$, $\Sigma_0 = 50 {\rm M_\odot} \pc^{-2}$ and we also have cut our disc so that $\rho=0$ for $R\geq 5\kpc$. This profile better matches the observed radial distribution of gas in the Galaxy \citep{KalberlaDedes2008,HeyerDame2015}. The initial density distribution is very smooth and we do not include any random noise. Despite this smoothness of the initial conditions, the gas flow in the bar region ends up being unsteady and turbulent because of the processes described in sect. 4 of \cite{Sormani+2018}.

\section{Results}
\label{sec:results}

\begin{figure*}
\includegraphics[width=1.0\textwidth]{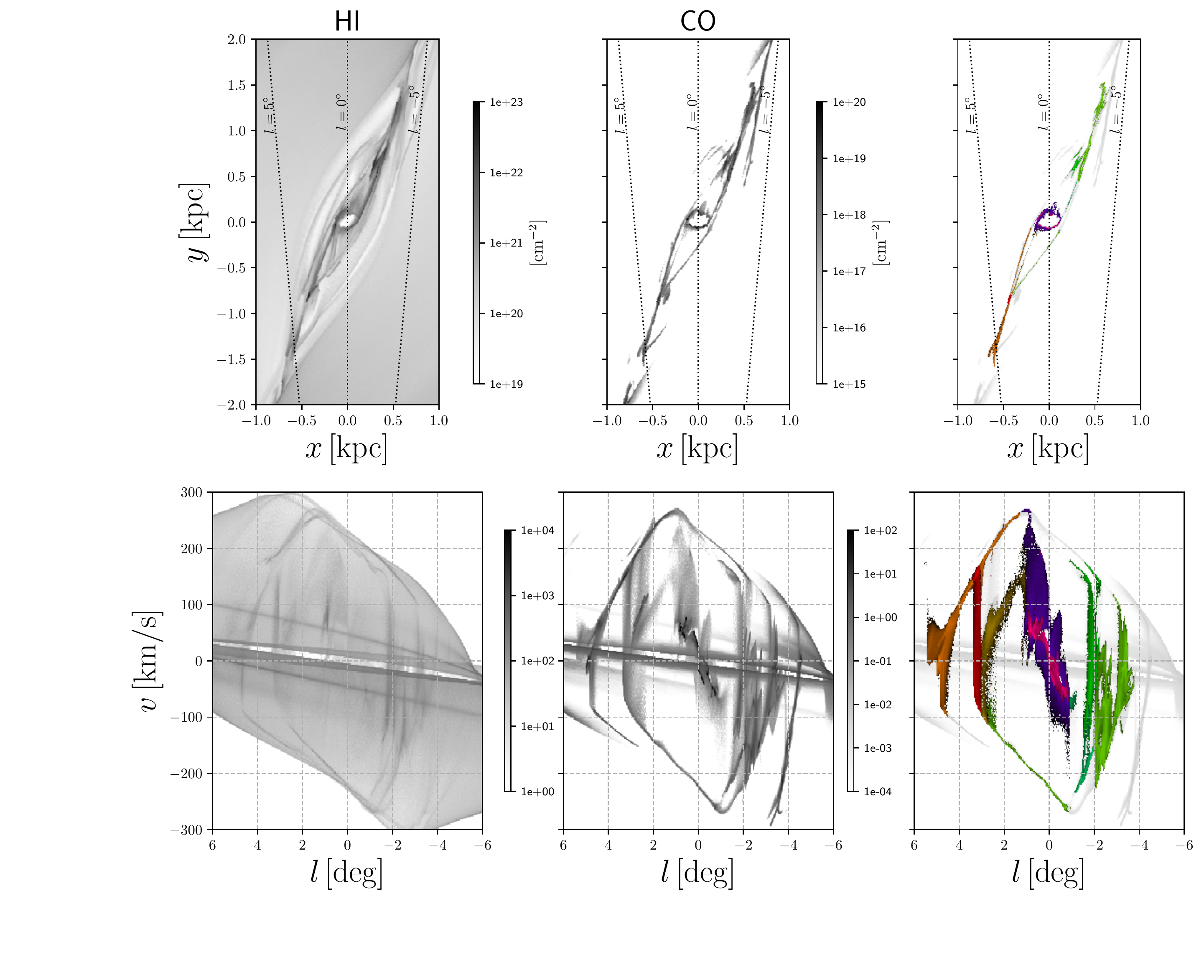}
\caption{The snapshot of our simulation at $t=178 \Myr$. \emph{Top row}: surface density of gas in the $(x,y)$ plane. \emph{Bottom row:} corresponding projections in the $(l,v)$ plane in the optically thin approximation and assuming that the angle between the Sun-GC line and the major axis of the bar is $\phi=20\degree$ \citep{BlandHawthornGerhard2016}. The left and middle column show HI and CO respectively as calculated by the chemical network included in the simulation. The right column shows a colour coded map on top of the CO emission, allowing one to identify corresponding structures in the $(x,y)$ and $(l,v)$ views. A movie showing a 3D visualisation of the snapshot shown in this figure is available at \url{http://www.ita.uni-heidelberg.de/~mattia/videos/EVF/flyby.mp4}.}
\label{fig:patches}
\end{figure*}

\begin{figure*}
\includegraphics[width=1.0\textwidth]{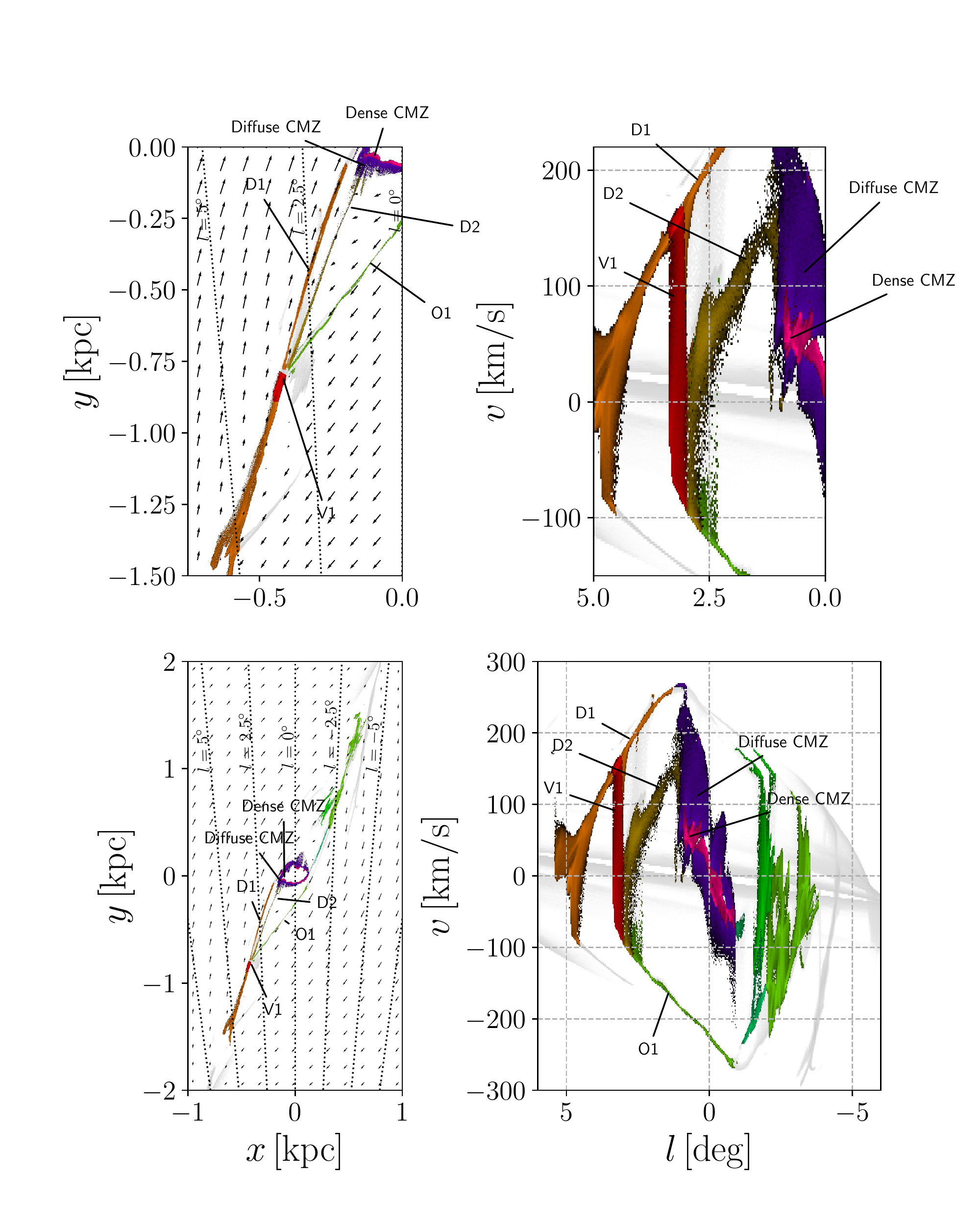}
\caption{Features in the $(x,y)$ plane and their projection to the $(l,v)$ plane for the simulation snapshot at $t=178 \Myr$. The top panels are zoom-ins of the bottom panels. Arrows in the left panels show the velocity field in the rotating frame of the bar. Labels mark some of the interesting features. The feature V1 resembles the EVF observed at $l=5.4\degree$ in Fig. \ref{fig:data}. The feature V1 originates as the material on the `overshooting' feature O1 (which has passed very close to the CMZ and brushed it) crashes onto the dust lane feature D1.}
\label{fig:patchesmap}
\end{figure*}

Figure \ref{fig:patches} shows a snapshot of our simulation at $t=178\Myr$. The top row shows the HI and CO surface density in the $(x,y)$ plane, while the bottom row shows the corresponding projections in the $(l,v)$ plane. To produce these projections, we bin each  {\sc Arepo} cell as a point in the $(l,v)$ plane with a weight proportional to the mass of the component of interest (HI or CO, as appropriate) and inversely proportional to the square of its distance from the Sun, as discussed in more detail in Section 3.6 of \citet{Sormani+2018}.
These projections assume that the gas is optically thin to HI and CO line emission, but accounting more accurately for line opacities would only change the strength of the emission and not its distribution in the $(l,v)$ plane. Figure \ref{fig:patchesmap} shows the correspondence between top down and projection plots in more detail, with labels that identify some of the interesting features.

Several features that resemble the observed EVFs can be identified in the various $(l,v)$ projections. A particularly prominent one is the red feature at $l\simeq 3 \degree$ labelled V1 in Fig. \ref{fig:patchesmap}. This feature has an extreme velocity dispersion ($\Delta v \sim 200\kms$) but is very localised in real $(x,y)$ space (it all originates from the small red patch visible in the top-left panel of Fig. \ref{fig:patchesmap}). This is precisely the main property that characterises the observed EVFs (see Sect. \ref{sec:observations}). The V1 feature connects the dust lane features D1 and D2 (see labels in Fig. \ref{fig:patchesmap}). This is remarkably similar to the what is observed for the EVF at $l\simeq 5.4 \degree$ in Fig. \ref{fig:data}, which connects the main observed dust lane L1 to the secondary dust lane L3. 

\begin{figure}
\includegraphics[width=0.5\textwidth]{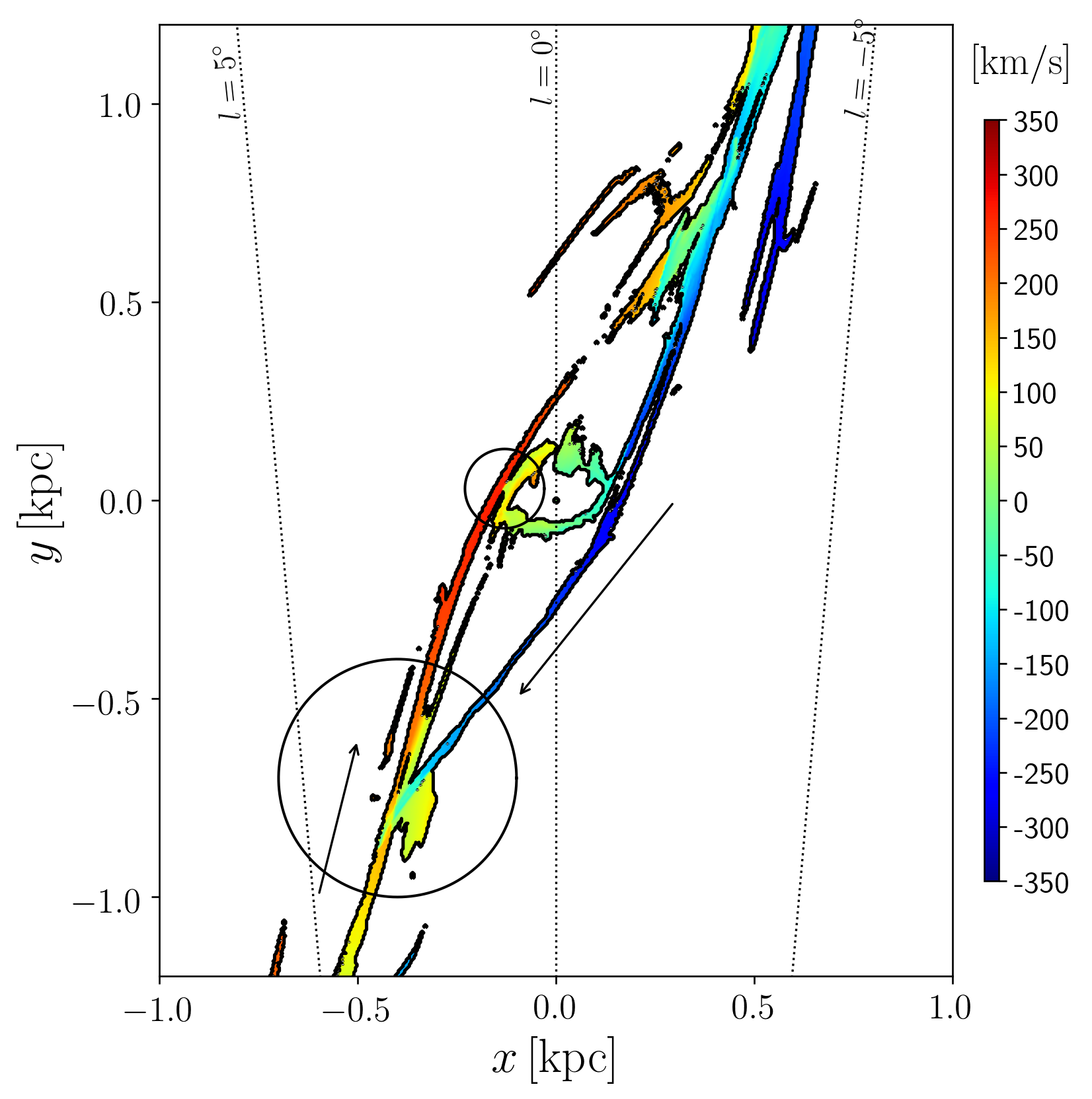}
\caption{Line-of-sight velocity in the $(x,y)$ plane. The larger circle highlights where the feature V1 shown in Fig. \ref{fig:patchesmap} originates. In this region, material with very different line-of-sight velocities collides, producing the large velocity dispersion observed in the $(l,v)$ plane. The smaller circle highlights a region at the outer edges of the CMZ, where the dust lane brushes the CMZ. This behaviour also brings into contact material with very different velocities and can give rise to EVFs.}
\label{fig:losvelocity}
\end{figure}

Inspection of the velocity fields in Figs. \ref{fig:patchesmap} and \ref{fig:losvelocity} reveals the origin of the feature V1. It originates as gas on the feature O1 crashes into the dust lane feature D1. The feature O1 is gas that has fallen along the dust lane on the opposite side, touched \& brushed the CMZ, and then continued its course until it crashed into the middle of feature D1.\footnote{Using high-sensitivity CO data \cite{MizunoFukui2004} have identified what might be the observational counterpart of the overshooting feature O1 (see crosses in their Fig. 3). This feature seems to connect to the $l=5.4\degree$ feature in the three-dimensional $(l,b,v)$ space in a manner very similar to how the O1 feature connects to the V1 feature in our simulations. This however needs to be confirmed by future observations.} When the feature O1 comes in contact with feature D1, the two have enormously different velocities. The signature of this collision in the $(l,v)$ plane is the extreme velocity dispersion that characterises the feature V1.

Figure \ref{fig:patchesmap} also shows the presence of several further features with high velocity dispersion at negative longitudes. These are coloured green. These features originate with a similar mechanism as the feature V1 discussed above. They look more crowded in the $(l,v)$ plane partly on account of projection effects (they are on the far side of the Galaxy). The production of the EVFs is a stochastic process in the simulation on account of the unsteady and turbulent flow that develops due to the processes described in section 4 of \cite{Sormani+2018}.

\begin{figure*}
\includegraphics[width=1.0\textwidth]{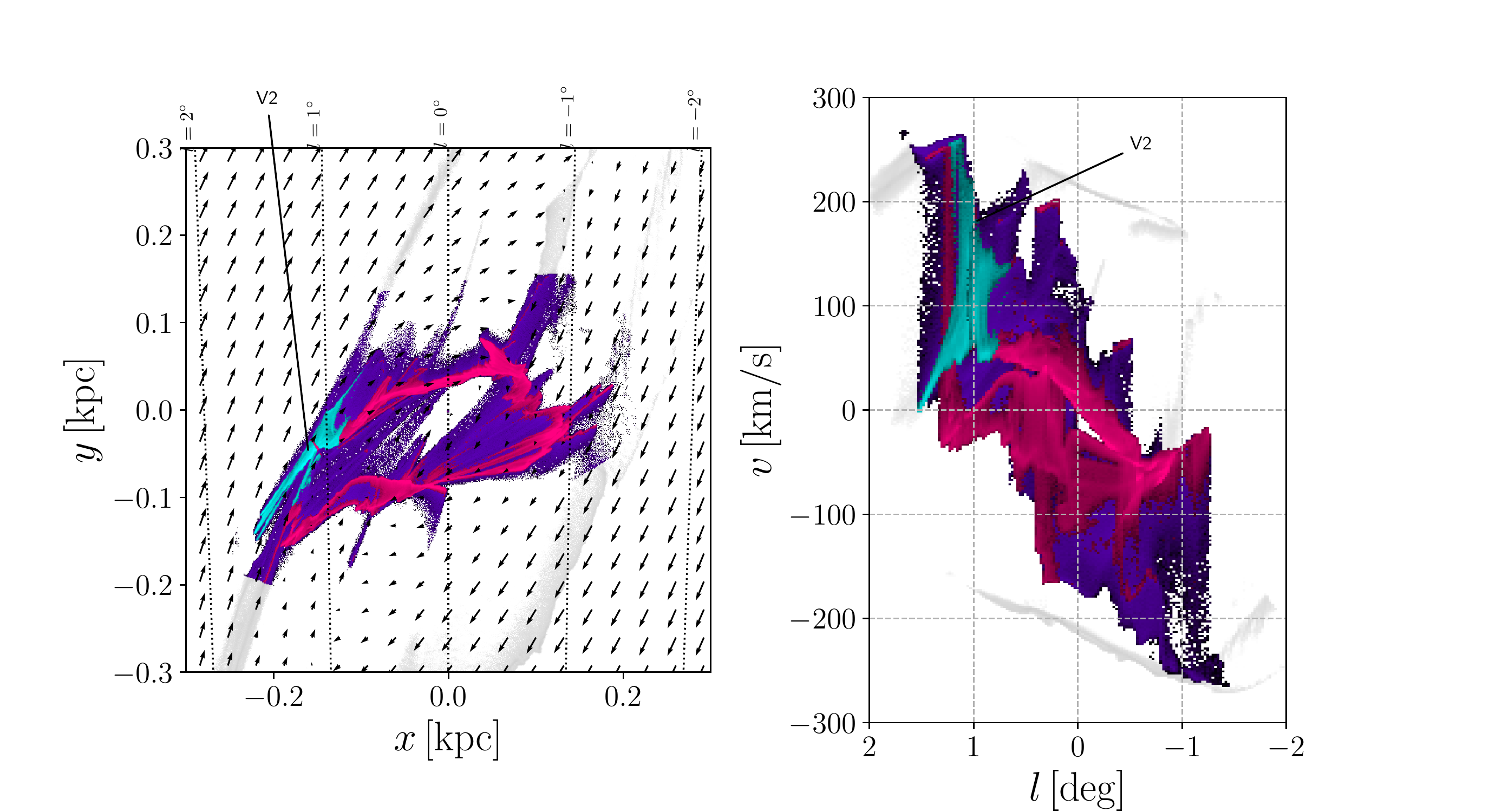}
\caption{Features in the $(x,y)$ plane and their projection to the $(l,v)$ plane in the central regions for the simulation snapshot at $t=191 \Myr$. The feature V2 illustrates the second type of EVF. This is created as incoming material from the dust lanes crashes into the CMZ.}
\label{fig:patchesmapCMZ}
\end{figure*}

A second type of broad-lined features that have a somewhat different origin than the ones described above also appear during the course of the simulation. Figure \ref{fig:patchesmapCMZ} shows an example of this second type of EVF. It is labelled V2 in the figure. This second type of feature originates as material that has been falling along the dust lanes crashes into the CMZ. The dense material in the CMZ typically has velocities much lower than those of the dust lanes, so when they collide they produce very broad-lined features like V2 in the figure. This feature has much in common with the observed EVF at $l=1.3\degree$ (see Fig. \ref{fig:data}). 

\begin{figure}
\includegraphics[width=0.45\textwidth]{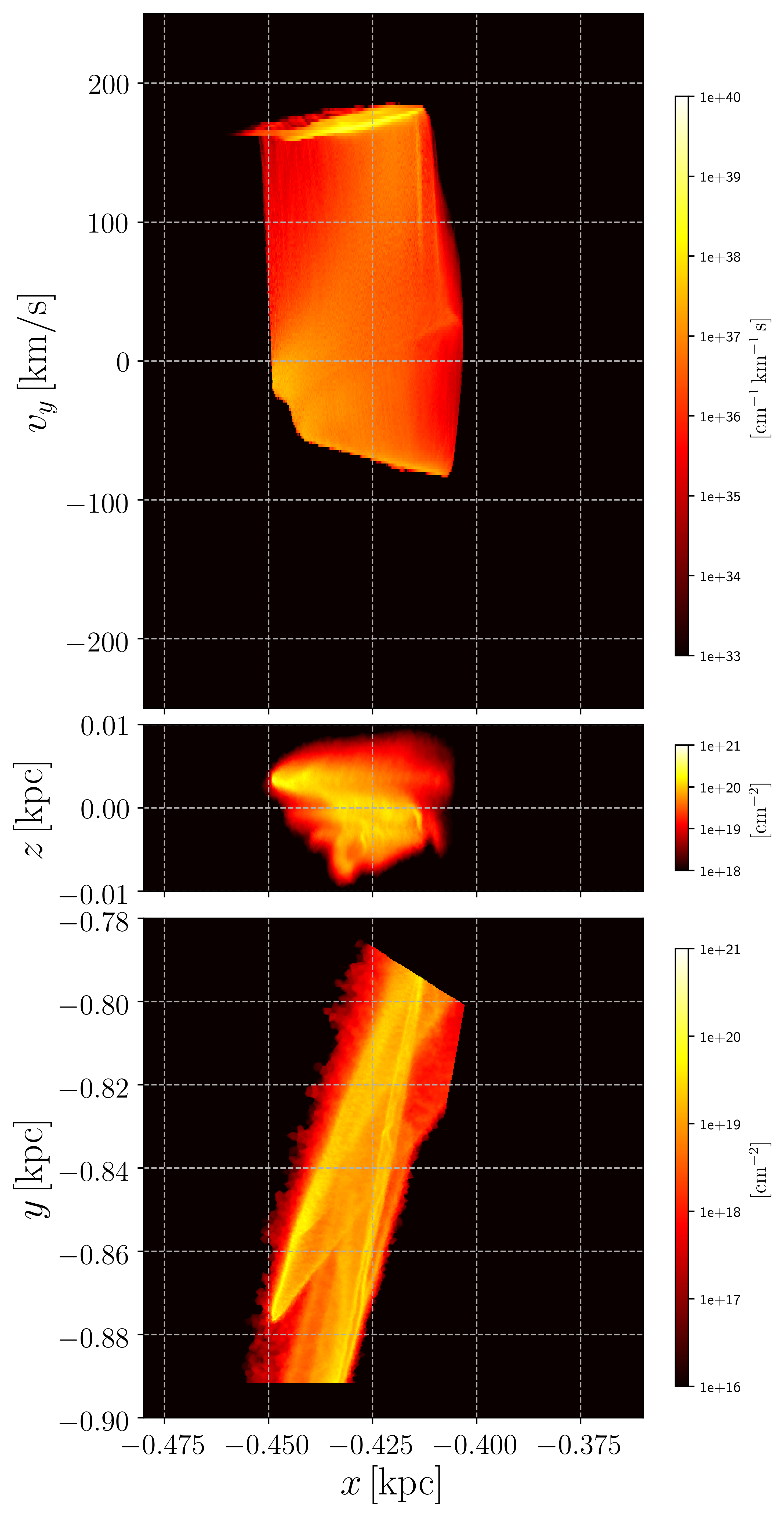}
\caption{Zoom-in that shows the 3D CO Position-Position-Velocity structure of the feature V1 in Fig. \ref{fig:patchesmap}. A movie showing the feature from different orientations is available at \url{http://www.ita.uni-heidelberg.de/~mattia/videos/EVF/zoomV1.mp4}.}
\label{fig:zoomV1}
\end{figure}

\begin{figure}
\includegraphics[width=0.45\textwidth]{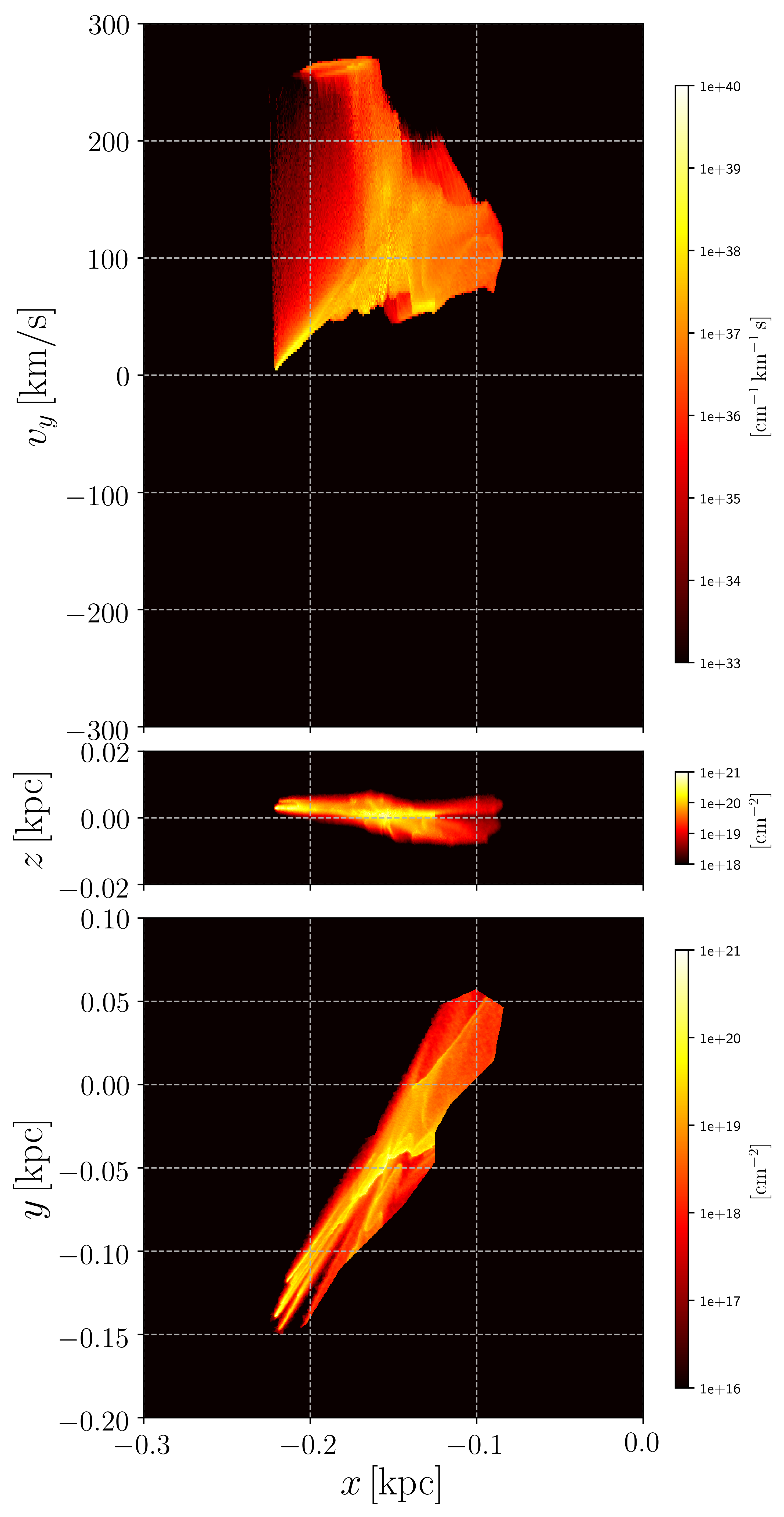}
\caption{Zoom-in that shows the 3D CO Position-Position-Velocity structure of the feature V2 in Fig. \ref{fig:patchesmapCMZ}. A movie showing the feature from different orientations is available at \url{http://www.ita.uni-heidelberg.de/~mattia/videos/EVF/zoomV2.mp4}.}
\label{fig:zoomV2}
\end{figure}

\section{Discussion}
\label{sec:discussion}

The results in the previous section suggest that at least some (perhaps most) of the EVFs found in the observations originate from collisions. These typically involve gas falling along the dust lanes that crashes with material with very different line-of-sight velocities. Our simulations show that this occurs naturally when gas flows in a barred potential and cannot be avoided: our initial conditions are prepared ensuring that the gas is as calm as possible (they are very smooth, symmetric and do not include any random noise), yet such collisions develop spontaneously. This happens even in the absence of any form of stellar feedback.\footnote{Indeed, the gas flow in a barred potential is inevitably unsteady and turbulent \citep[][]{Sormani+2018}. This is well illustrated for example in the top-left panel in Fig. \ref{fig:patches}, which shows that inside the bar region the flow is structured and unsteady, in striking contrast with the flow just outside the bar region which is extremely smooth and steady.} 

The extended velocity features occur frequently in the simulations although perhaps at any given time there are somewhat fewer of them in a synthetic $(l,v)$ diagram than in its observational counterpart. This is probably a consequence of the fact that we have tried to keep the gas flow as smooth as possible, while in the real Galaxy more collisions should be expected on account of the facts that the initial conditions are most likely not smooth and that additional processes contribute to produce more unsteadiness and turbulence (stellar feedback, perturbations from satellite galaxies that punch through the MW disc, etc). Thus our simulations provide a lower limit on the number of EVF-producing collisions that might be expected in the real Galaxy.

Our interpretation naturally explains most of the key observational properties listed in Section~\ref{sec:observations}. Property (i) is satisfied because this is the property by which we select features in the simulation to compare to the observed EVFs. Property (ii) is satisfied because the collision sites have limited extension in real $(x,y)$ space, so the features are localised in the $(l,b)$ plane.  Property (iv) is satisfied because collisions in the simulations happen preferentially in the two quadrants $(l>0,v>0)$ and $(l<0,v<0)$, although not exclusively (see for example the green material in Fig. \ref{fig:patchesmap}). Property (v) is satisfied because colliding clouds are part of the general large-scale flow and so their velocities are always within the limits defined by the TVC. Property (vi) is satisfied because in our interpretation some of the features are naturally connected with the dust lanes. 

With our existing simulation, we are not able to verify whether features formed in this way satisfy Property (iii). One of the unrealistic properties of our simulation is that the gas layer is too thin compared to observations (typical thickness of molecular gas in the simulations is only $H \sim 10 \pc$), probably due to the lack of stellar feedback \citep[see the discussion in section 5.5.1 of][]{Sormani+2018}. The thinness of the simulations can also be appreciated from the movies linked in the Supplementary Information section below. Hence, on scales much larger than $\sim10\pc$ the gas is always more elongated in longitude than in latitude in our simulations, contrary to Property (iii). However, we might argue that both Property (iii) and (vii) may be expected for more realistic (and therefore more vertically `puffed up') clouds within the context of our interpretation. When two clouds collide at high speed, we expect them to be compressed in the direction of motion (in this case, the $l$ direction). This might explain Property (iii). Similarly, one might expect that a collision produces a strong compression shock on one side, visible as a sharp edge (Property vii). 

The masses of the features in the simulations are comparable to the masses of the observed EVFs. For example. the mass of the feature V1 in the simulation is $\simeq 2.5\times 10^{6} \, \rm M_\odot$ while the mass of the observed $l=5.4\degree$ feature has been estimated by \cite{Liszt2006} as $\simeq 5 \times 10^{6} \, \rm M_\odot$. This is a good agreement given that (a) the processes that produce the EVFs and therefore their masses are stochastic and (b) the masses measured from the observations are very uncertain due to the uncertainty in the CO-to-H$_{2}$ conversion factor ($X_{\rm CO}$). Indeed, standard assumptions made to calibrate $X_{\rm CO}$ such as virial equilibrium \citep[e.g.][]{Bolatto+2013} are most likely not valid for the features considered here which are in a highly dynamical environment.

As noted in Sect. \ref{sec:observations} (Property viii), observed EVFs typically have a very complicated internal Position-Position-Velocity (PPV) structure and break-up into several sub-components with strong velocity gradients when observed at very high resolution. What is the small scale structure of the EVFs obtained in the simulations? To investigate this we show in Figs. \ref{fig:zoomV1} and \ref{fig:zoomV2} the CO PPV maps of the features V1 and V2 studied above. Movies that show the same features from different orientations are also available at the link provided in the Supplementary Information section below. These show that V1 and V2 are indeed connected structures in 3D physical Position-Position-Position space, and not coincidental amalgamations of unconnected components. The simulated EVFs posses a certain degree of internal structure (particularly the kinematical structure of V2 appears to be significantly more complex than of V1), but the real observed EVFs display a much higher degree of complexity (compare Figs. \ref{fig:zoomV1} and \ref{fig:zoomV2} with figs. 6,7,8,9 of \citealt{Liszt2006} and fig. 30 of \citealt{Longmore+2017}). This is not unexpected given that our simulations start out very smooth and lack any kind of stellar feedback, self-gravity and/or initial noise that could generate substructure, so that on small scales clouds tend be much smoother than their counterparts in the real Galaxy. It is however interesting to note that the simulated EVFs \emph{do} have some substructure due to the unsteady gas flow caused by the bar, in contrast to the gas outside the bar region which is extremely smooth.\footnote{The smoothness outside the bar region can be appreciated for example in the visualisation at the following link: \url{http://www.ita.uni-heidelberg.de/~mattia/videos/EVF/flyby.mp4}.}\footnote{The observations, as seen for example in \cite{Longmore+2017}, also seem to indicate a possible connection between the statistics of the small-scale velocity structure and the type of EVF. Since our simulations are currently unable to reproduce the small-scale complexity of EVFs, we refrain from specifying the expected statistics of the different small-scale velocity structures. However, this is worth further study.} Another aspect that is evident from Figs. \ref{fig:zoomV1} and \ref{fig:zoomV2} is the small vertical extent (i.e. in the $z$ direction) of our simulations discussed above. Despite these caveats, the comparison shows that the simulated EVFs may be identified with the bulk gas of the observed EVFs. 

Finally, we note the following. In the previous section we have identified two mechanisms that produce collisions (and therefore EVFs) in our simulations. The first is overshooting material which collides with the dust lanes on the opposite side, exemplified by feature V1 in Fig. \ref{fig:patchesmap}. The second is material on the dust lanes which collides with CMZ material, exemplified by feature V2 in Fig. \ref{fig:patchesmapCMZ}. However we cannot exclude that in a more turbulent, realistic environment further mechanisms that generate collisions are possible. For example, multiple dust lanes are generally expected to be very close in real space although they have very different line-of-sight velocities. A relatively small perturbation to the velocity field (induced for example by an external perturbation such as accretion from the circumgalactic medium or stellar feedback) may cause them to touch. This would lead to a transfer of material between the two (the faster dust lane is decelerated, while the slower one is accelerated), which in the $(l,v)$ diagram would show up as an EVF. The key point is that velocity dispersions of the order of $\sim 100 \mhyphen 200 \kms$ (comparable to the velocity of the Sun around the Galactic centre) suggest that collisions between large-scale Galactic flows are involved. The presence of a bar creates the perfect environment to make such collisions likely.

\subsection{Comparison with previous work}

Compared to the other interpretations (a) to (d) listed in the introduction, we note the following. Unlike interpretation (a), according to which EVFs are extended structures that coincidentally line parallel to the line-of-sight, our interpretation does not suffer from the `finger of god' effect. The patches of gas producing the EVFs in our simulations are always localised in $(x,y)$ space and in general do not correspond to structures which are elongated along the line-of-sight. For example, we have verified that our features remain `extended' in the $(l,v)$ plane even if observed at different angles $\phi$ between the major axis of the bar and the Sun-GC line. 

According to interpretation (b) magnetic instabilities alone (without a bar potential) are responsible for creating the EVFs. However, the synthetic $(l,v)$ diagrams produced from simulations of this mechanism performed to date  \citep{Machida+2009,Suzuki+2015,Kakiuchi+2018} do not seem to be able to convincingly reproduce the morphology of the EVFs in the $(l,v)$ plane (Properties i and ii in Sect. \ref{sec:observations}). Moreover, in this interpretation the connection with the dust lanes of the MW bar (Property vi in Sect. \ref{sec:observations}) remains unexplained. Nevertheless, it is possible that magnetic fields, when added on top of the bar potential, play a role in shaping the properties and morphologies of the EVFs. 

Interpretation (c) assumes that EVFs are created by gravitational kicks around IMBHs. According to this interpretation, the large velocity dispersion seen in an EVF should depend on the impact parameter of the incoming gas cloud relative to the IMBH and on the mass of the IMBH, and should have no relation to the TVC and/or to the dust lanes features of the MW. Hence in this interpretation it is unclear why the EVFs never extend beyond the TVC at their longitudes (Property v) and why they seem to be associated with the dust lanes of the MW (Property vi). This interpretation also posits an ad-hoc assumption, namely the presence of IMBHs, which is unnecessary since it can be avoided in our interpretation. Finally, we note that in the case of the CO-0.40-0.22 cloud, an EVF that has been claimed to be the signature of an IMBH close to the Galactic centre \citep{Oka+2017}, constraints on the radio spectrum and a detection of a mid-infrared point source both disfavour the presence of an IMBH \citep{Ravi+2018}.

The interpretation (d) of \cite{Fux1999} is essentially the same that we have given in this paper, but in an embryonic state.  The simulations of \cite{Fux1999} did not possess the necessary resolution to actually see the EVFs in the synthetic $(l,v)$ diagrams. Fux speculated about the implications of his simulations and imagined what he would have seen if he had higher resolution. We have refined the \cite{Fux1999} interpretation by correcting some parts (e.g.\ the clumps are not really `crossing' the dust lane and exiting on the other side as Fux suggested, but instead are joining and merging with the dust lane and then flowing together towards the central regions) and filling in some details (e.g.\ the origin of some of the clumps hitting the dust lane are clumps from the dust lanes on the other side that have overshot).

\begin{figure}
\includegraphics[width=0.4\textwidth]{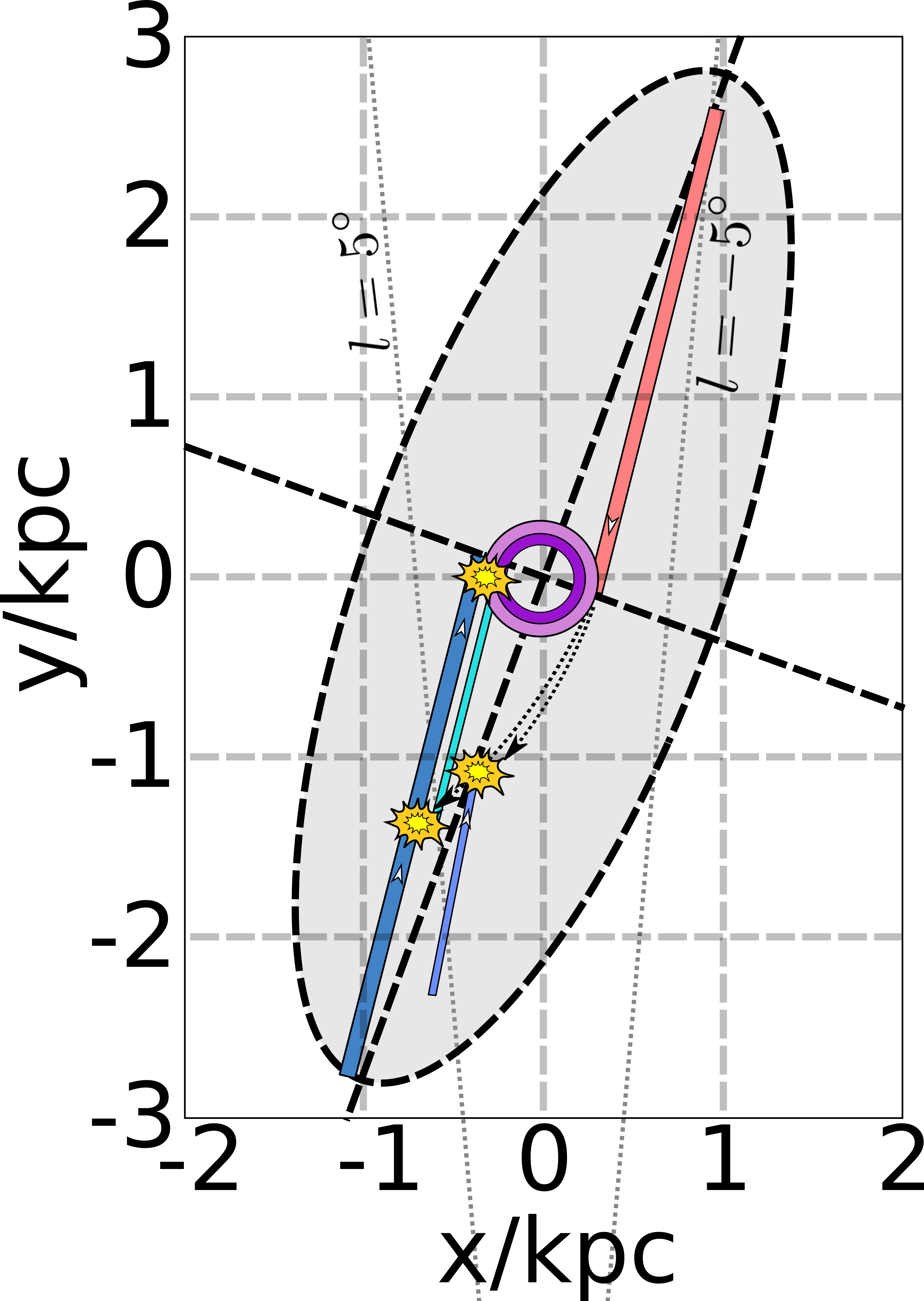}
\caption{Sketch of how the geometry of the gas surrounding the CMZ might look like according to our interpretation. Coloured straight lines represent the various dust lanes of the MW. The purple circle represents the CMZ. The two yellow clouds on the near side dust lanes represent the $l=5.4\degree$ and $l=3.2\degree$ (aka Bania Clump 2) EVFs in Fig. \ref{fig:data} respectively. The yellow cloud on at the intersection between dust lanes and CMZ represents the $l=1.3\degree$ EVF in Fig. \ref{fig:data}.}
\label{fig:sketch}
\end{figure}

\subsection{Implications for the observations} \label{sec:implications}

In Section~\ref{sec:results}, we showed that we can distinguish two basic types of EVFs in the simulation. The first is produced by material on the dust lanes that collides with overshooting material from the other side. The second is produced by material on the dust lanes which collides with CMZ material. 

In the observations some features can be associated quite clearly with one or the other of these two possibilities. For example, the features at $l=5.4\degree$ and $l=3.2\degree$ in Fig. \ref{fig:data} are most likely of the first type (because of the way they are associated with the dust lane features L1-L3 and because their longitudes place them outside the CMZ), while the feature at $l=1.3\degree$ is most likely of the second type (because it connects with dense CMZ gas, see Fig \ref{fig:data}). Figure \ref{fig:sketch} shows a sketch of how the geometry of the gas surrounding the CMZ might look like according to our interpretation. For other features in the observations the situation is more ambiguous, and one needs to study this on a case by case basis, checking for example the connection between them and other features such as dust lane features and using high resolution data, which is outside the scope of the present paper.

The HI projection in Fig. \ref{fig:patches} displays many features that cannot be seen in the CO projection. Thus we expect that several features which are invisible in high-density tracers may be detected in low-density observational tracers such as HI (or the CO $J=1\to0$ line observed with high enough sensitivity). The features identified in low-density tracers can be used to connect the features seen in higher-density tracers such as NH$_3$ or HCN. This will be necessary to get a complete picture of the 3D geometry and gas flows in and around the CMZ.

Finally, it is worth noting that if our interpretation of the $l=5.4\degree$ and $l=3.2\degree$ EVFs is correct, we are directly witnessing collisions at a relative speed of $\Delta v\sim 200\kms$. This is a perfect laboratory for studying what happens when two molecular clouds with masses in excess of $M=10^6 \, \rm M_\odot$ collide with each other with extreme velocities. We expect to find a rich chemistry and the presence of shock tracers associated with these features in the observations. If the interpretation of the $l=1.3\degree$ EVF is correct, we are directly witnessing material that is accreting onto the CMZ. Studying this feature in more detail can, therefore, give insight on the process of accretion as it is happening and on the physical and chemical condition of the accreted gas.

\section{Summary} \label{sec:conclusion}

Surrounding the Galactic centre there exist an enigmatic population of compact molecular clouds with extreme velocity dispersions. These Extended Velocity Features (EVFs) dominate the kinematics of gas just outside the Central Molecular Zone. We have used hydrodynamical simulations of gas flow in a barred potential to interpret these clouds. We have found that similar features occur naturally in these simulations. They originate from collisions between material that is falling along the dust lanes of the bar and material with substantially different line-of-sight velocities. We have distinguished between two types:
\begin{enumerate}
\item EVFs like the feature V1 in Fig. \ref{fig:patchesmap} which originate from the collision between material on the dust lanes and material that has `overshot' from the dust lane on the opposite side;
\item EVFs like the feature V2 in Fig. \ref{fig:patchesmapCMZ} which originate from the collision between material on the dust lanes and material belonging to the CMZ.
\end{enumerate}
Examples of both types of features can be identified in the observations. The sketch in Fig. \ref{fig:sketch} shows our proposed interpretation of the most unambiguous features. Other features can be identified in the data, but the interpretation is more ambiguous and will require more careful analysis with higher resolution observations.

If our interpretation is correct, we are witnessing clouds colliding at relative velocities of $\Delta v \sim 200\kms$ (e.g. the $l=5.4\degree$ and $l=3.2\degree$ clouds). This provides an excellent laboratory to study extreme cloud collisions. We are also directly witnessing gas being accreted onto the CMZ (e.g. the $l=1.3\degree$ cloud). This provides a unique opportunity to study how gas is accreted and the physical and chemical properties of the accreted gas.

\section*{Supplementary information}

Movies showing the time evolution of the simulations, a 3D visualisation of the snapshot shown in Fig. \ref{fig:patchesmap} and the three dimensional PPV structure of the features V1 and V2 can be found at the following link: \url{http://www.ita.uni-heidelberg.de/~mattia/download.html}.

\section*{Acknowledgements}

MCS thanks Tom Dame and Harvey Liszt for insightful comments and discussions. We thank Jean-charles Lambert for developing {\sc GLNEMO2}, a freely-distributed interactive visualization 3D software for N-body snapshots which is publicly available at the following link: \url{https://projets.lam.fr/projects/glnemo2}. MCS, RGT, SCOG, and RSK acknowledge support from the Deutsche Forschungsgemeinschaft via the Collaborative Research Centre (SFB 881) ``The Milky Way System'' (subprojects B1, B2, and B8) and the Priority Program SPP 1573 ``Physics of the Interstellar Medium'' (grant numbers KL 1358/18.1, KL 1358/19.2, and GL 668/2-1). RSK furthermore thanks the European Research Council for funding in the ERC Advanced Grant STARLIGHT (project number 339177). ATB would like to acknowledge the funding provided from the European Union's Horizon 2020 research and innovation programme (grant agreement No 726384). CDB and HPC gratefully acknowledges support for this work from the National Science Foundation under Grant No. (1816715). PCC acknowledges support from the Science and Technology Facilities Council (under grant ST/N00706/1) and StarFormMapper, a project that has received funding from the European Union's Horizon 2020 Research and Innovation Programme, under grant agreement no. 687528. HPH thanks the LSSTC Data Science Fellowship Program, which is funded by LSSTC, NSF Cybertraining Grant No. 1829740, the Brinson Foundation, and the Moore Foundation; his participation in the program has benefited this work. RJS gratefully acknowledges an STFC Ernest Rutherford fellowship (grant ST/N00485X/1) and HPC from the Durham DiRAC supercomputing facility (grants ST/P002293/1, ST/R002371/1, ST/S002502/1, and ST/R000832/1). The authors acknowledge support by the state of Baden-W\"urttemberg through bwHPC and the German Research Foundation (DFG) through grant INST 35/1134-1 FUGG.

%%%%%%%%%%%%%%%%%%%%%%%%%%%%%%%%%%%%%%%%%
\def\aap{A\&A}\def\aj{AJ}\def\apj{ApJ}\def\mnras{MNRAS}\def\araa{ARA\&A}\def\aapr{Astronomy \&
 Astrophysics Review}\def\apjs{ApJS}\def\apjl{ApJ}\def\pasj{PASJ}\def\nat{Nature}\def\prd{Phys. Rev. D}
\def\ssr{Space Sci. Rev.}\def\pasp{PASP}\def\aaps{A\&AS}
\bibliographystyle{mn2e}
\bibliography{bibliography}

\appendix

\section{Gravitational potential} \label{sec:potential}

\begin{figure}
\includegraphics[width=84mm]{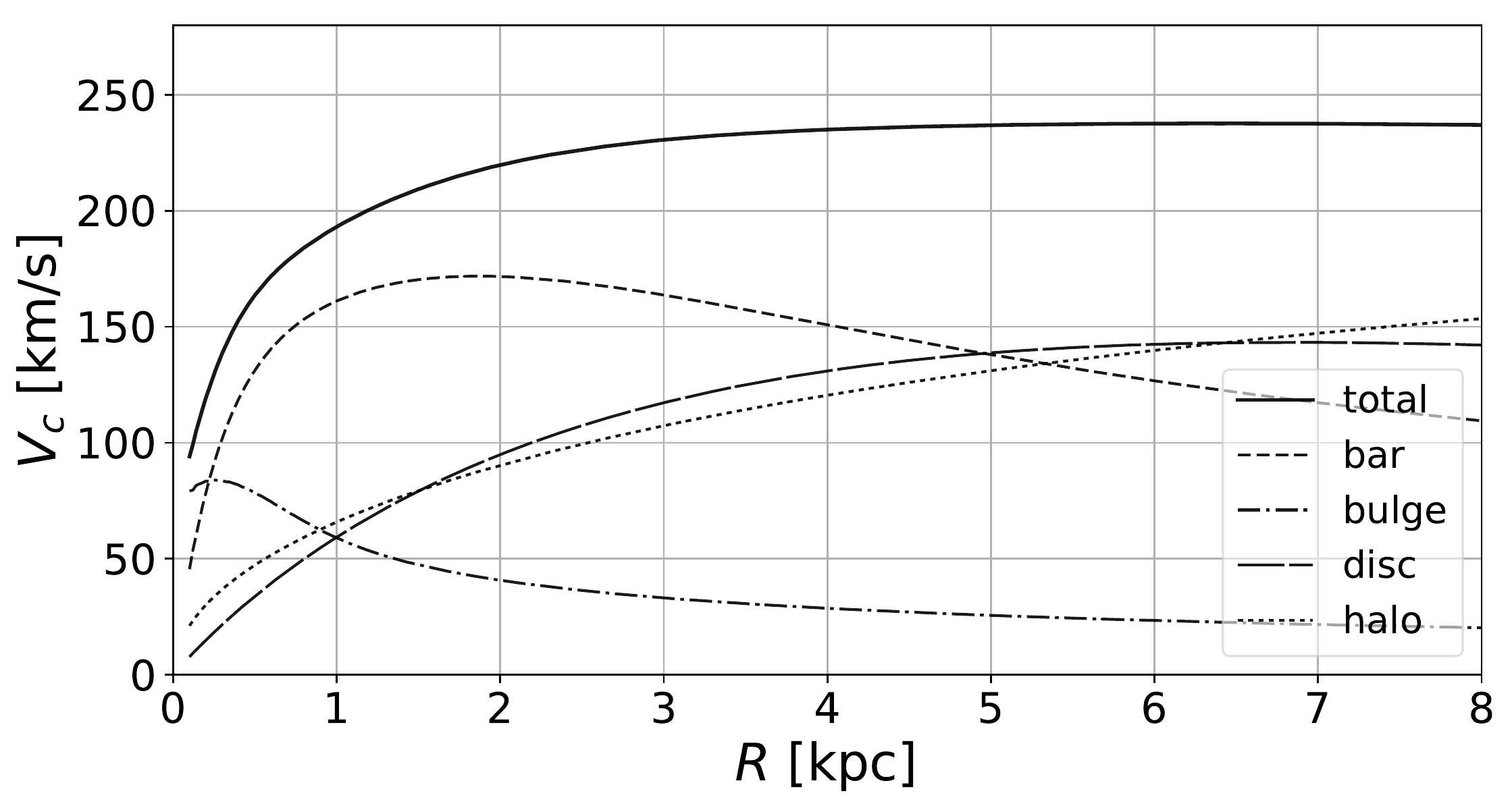}
\caption{The circular velocity curve in the plane $z=0$ for the potential used in this paper. The separate contributions from bar, bulge, disc and halo are also shown.}
\label{fig:vc}
\end{figure}

\begin{figure}
\includegraphics[width=84mm]{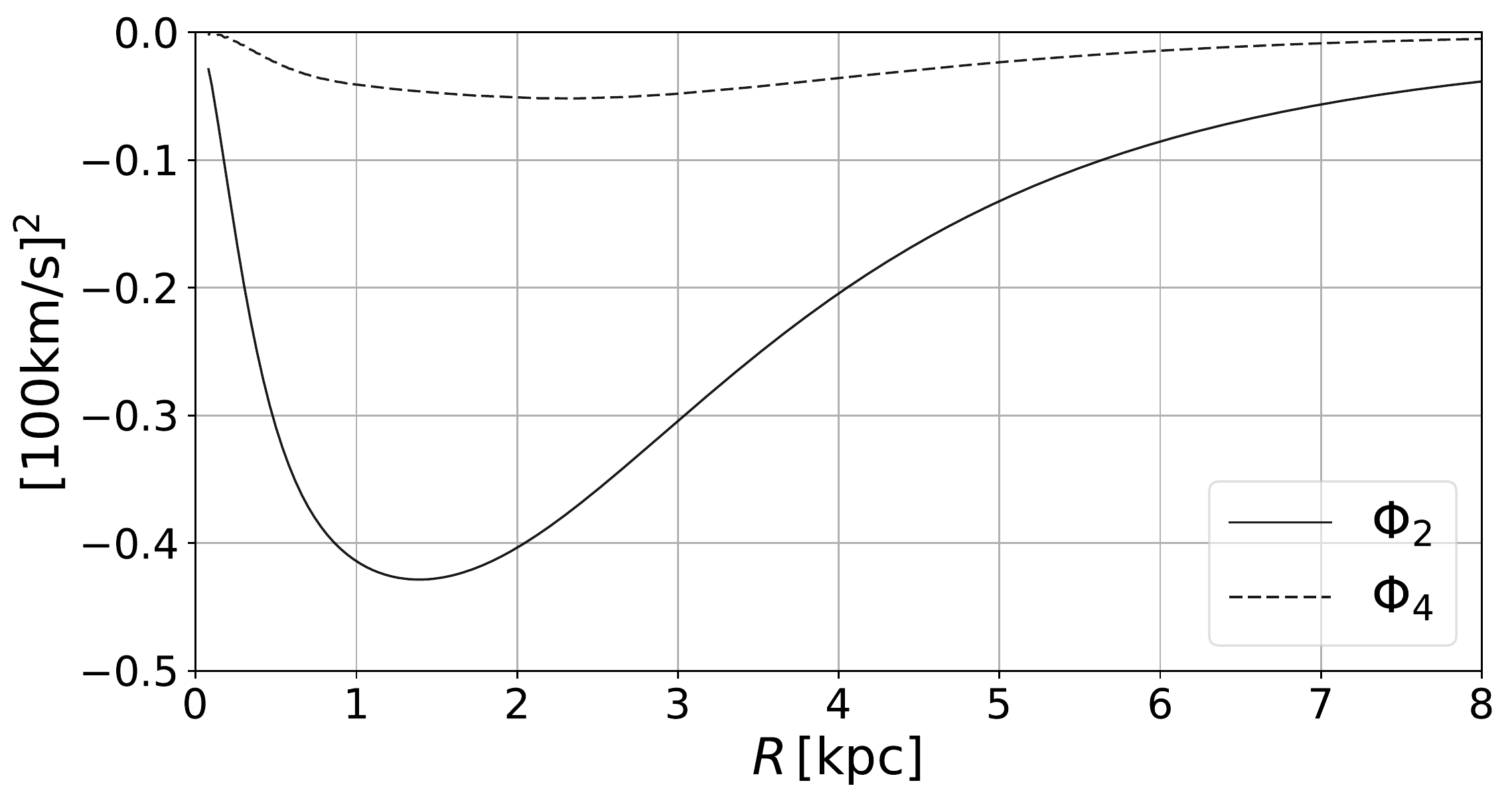}
\caption{The quadrupole $\Phi_2$ and octupole $\Phi_4$  in the plane $z=0$ for the potential used in this paper. These are defined by the multipole expansion of the potential in the plane of the Galaxy, $\Phi(R,\phi) = \Phi_0(R) + \sum_{m=1}^{\infty} \Phi_m(R) \cos\left(m \phi + \phi_m\right)$ where $\phi_m$ are constants and $\{R,\phi,z\}$ denote standard cylindrical  coordinates.}
\label{fig:multipoles}
\end{figure}

We employ a realistic external gravitational potential that is the sum of four components: bar, bulge, disc, and halo. The axisymmetric part is derived from the work of \cite{McMillan2017}, whose potential is created to fit observational constraints and to be consistent with expectations from theoretical modelling of the Milky Way as a whole. The bar and the bulge are built to be consistent with observational constraints from near-infrared photometry \citep{Launhardt+2002} and with dynamical constraints on the quadrupole of the bar (\citealt{SBM2015c}; see also \citealt{Ridley+2017}). The bar rotates with a constant pattern speed of $\Omega_{\rm p}=40\kms\, {\rm kpc^{-1}}$. The axisymmetric part (the velocity curve) and the first few multipoles are shown in Figs. \ref{fig:vc} and \ref{fig:multipoles}. The details of each component of the potential are as follows.

\subparagraph{Bulge.} This component is generated by the following density distribution:
\begin{equation}
\rho_{\rm b} = \frac{ \rho_{{\rm b}0} }{(1 + a/a_0)^\alpha} \exp\left[ - \left( a/a_{\rm cut}\right)^2 \right]
\end{equation}
where
\begin{equation}
a = \sqrt{x^2 + y^2 + \frac{z^2}{q_{\rm b}^2}}.
\end{equation}
We use the following parameters: $\alpha=1.8$, $\rho_{{\rm b}0} = 9.5 \times 10^4 \,\rm  M_\odot \pc^{-3}$, $a_{\rm cut}=0.5\kpc$, $q_{\rm b}=0.5$ and $a_0= 10^{-3} \kpc$.

\subparagraph{Bar.} The density of the bar is taken to be:
\begin{equation}
\rho_{\rm B} = \rho_{{\rm B}1}  \exp\left( - a_1/a_{\rm B1} \right) + \rho_{{\rm B}2}  \exp\left( - a_2/a_{\rm B2} \right),
\end{equation}
where
\begin{align}
a_1 & =\sqrt{x^2 + \frac{y^2+z^2}{q_{\rm B1}^2} }, \\
a_2 & =\sqrt{x^2 + \frac{y^2+z^2}{q_{\rm B2}^2} }.
\end{align}
We use the following values for the parameters: $\rho_{{\rm B}1}=16 \,\rm M_\odot \pc^{-3}$, $a_{\rm B1}=0.3\kpc$, $q_{\rm B1}=0.5$, $\rho_{{\rm B}2}=3 \,\rm M_\odot \pc^{-3}$, $a_{\rm B2}=1\kpc$ and $q_{\rm B2}=0.5$.

\subparagraph{Disc.} The disc is the sum of a thick and a thin disc \citep{GilmoreReid1983}. The density distribution is:
\begin{equation}
\rho_{\rm d} = \frac{\Sigma_1}{2 z_1} \exp \left( -\frac{|z|}{z_1} - \frac{R}{R_{{\rm d} 1}} \right) + \frac{\Sigma_2}{2 z_2} \exp \left( -\frac{|z|}{z_2} - \frac{R}{R_{{\rm d} 2}} \right),
\end{equation}
where $\Sigma_1 = 572\, \rm M_\odot \kpc^{-2}$, $R_{\rm{d}1} = 2.9\kpc$, $z_1=0.3\kpc$, $\Sigma_2 = 147\, \rm M_\odot \kpc^{-2}$, $R_{\rm{d}2}=3.31\kpc$ and $z_2=0.9\kpc$.

\subparagraph{Halo.} This is a simple \cite{NFW96} profile. The density distribution is:
\begin{equation}
\rho_{\rm h} = \frac{\rho_{{\rm h}0}}{x (1 + x)^2}
\end{equation}
where $x = r/r_{\rm h}$, $r=(x^2+y^2+z^2)^{1/2}$, $\rho_{\rm{h}0}=0.00846 \, \rm M_\odot \pc^{-3}$, and $r_{\rm h} = 20.2\kpc$.

\end{document}